\documentclass[a4paper,12pt]{article}
\usepackage{amsmath}
\usepackage{amsfonts}
\usepackage{amssymb}
\usepackage{amstext}
\usepackage{graphicx}
\def\slashi#1{\rlap{\sl/}#1}

\begin{document}

\begin{titlepage}

\rightline{UAB-FT-663}
\rightline{PRL-TH/AP-09/1}
\vskip 2.5cm
\begin{center}
{\Large {\bf  Dipolar Dark Matter}}
\vskip 2cm
{\large Eduard Mass\'{o}$^1$, Subhendra Mohanty$^2$ and Soumya
Rao$^2$}\\
\vskip 0.5cm
$^1${\it Grup de F{\'\i}sica Te{\`o}rica and Institut de F{\'\i}sica
d'Altes Energies\\
Universitat Aut\`{o}noma de Barcelona, 08193 Bellaterra,
Spain}\\[0.2cm]
$^2${\it Physical Research Laboratory, Ahmedabad 380009,
India}\\
\vskip 1cm
\begin{abstract}

If dark matter (DM) has non-zero direct or transition, electric or
magnetic dipole moment then it can scatter nucleons
electromagnetically in direct detection experiments. Using the
results from experiments like XENON, CDMS, DAMA and COGENT we put
bounds on the electric and magnetic dipole moments of DM. If DM
consists of Dirac fermions with direct dipole moments, then DM of mass
less than 10 GeV is consistent with the DAMA signal and with null
results of other experiments. If on the other hand DM consists of
Majorana fermions then they can  have only non-zero transition moments
between different mass eigenstates.  We find that Majorana fermions
with mass  $m_\chi >38$ GeV and mass splitting  of the
order of (50-200) keV  can explain   the DAMA signal and the
null
observations from other experiments and in addition give the observed relic density
of DM by dipole-mediated annihilation. This parameter space for the mass and  for dipole moments is allowed by limits from L3 but may have observable signals at LHC.

\end{abstract}
\end{center}
\end{titlepage}

\newpage
\section{Introduction}\label{intro}

Experimental observations mainly of dynamics of spiral galaxies
and galaxy clusters indicate the existence of dark matter
(DM).  Cosmological observations confirm the existence of
DM and in addition show that the bulk of it must be non-baryonic
\cite{Drees}.
In this paper we consider weakly interacting massive particles (WIMPs)
as candidates for DM, but
we adopt a model independent and
phenomenological approach.

The mass and cross section of the DM (which is expected to have a
local density of about 0.3 GeV/cm$^3$ and velocity w.r.t the Earth
of about 200 km/sec \cite{gju96}) is probed by direct detection
experiments like XENON \cite{xenon}, CDMS \cite{cdms}, DAMA
\cite{dama} and COGENT \cite{cogent}. These experiments detect DM
scattering off nuclei by measuring the recoil energy of the nuclei.
The energy threshold of such detectors is typically of the order of
a few keV. Of these experiments, the DAMA experiment observed an
annual modulation in its signal which could have been due to DM
scattering.  However none of the other experiments conducting direct
DM searches have seen evidence of such an event. Theoretically if
one considers spin-independent interaction
\cite{gelmini,Bottino:2008mf} of these WIMPs with the nuclei then it
is found that only in the low mass range from about 5 to 10 GeV one
can reconcile DAMA signal with the null results from other
experiments.  There have been proposals of inelastic scattering of
DM \cite{weiner} by nuclei which results in heavier DM mass being
allowed for explaining all the existing data.

Although DM has zero electric charge it may
couple to photons through loops in the form of electric and magnetic
dipole moments.
Here we study WIMPs which are endowed with such dipole moments
and thus can interact feebly via electromagnetic interaction
\cite{mpos00,ksig04}.

We first will consider the case that DM is a Dirac
fermion.  The effective Lagrangian for coupling of a Dirac fermion
$\chi$ having an electric dipole moment $\mathcal{D}$  and a magnetic
dipole moment $\mu$  to a electromagnetic field $\mathcal{F}^{\mu\nu}$
is
\begin{equation}
\mathcal{L}_{elast}=-\frac{i}{2}\,
\bar{\chi}\sigma_{\mu\nu}(\mu+\gamma_5\mathcal{D})\chi\mathcal{F}^{\mu\nu}.
\label{L_elastic}
\end{equation}

\begin{figure}
\begin{center}
\includegraphics[width=5cm,height=3cm]{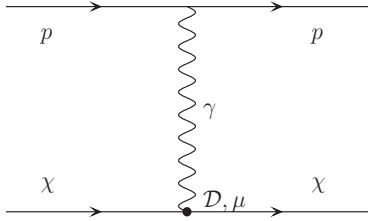}
\caption{Electromagnetic scattering of a proton with DM with non-zero
dipole moments.}
\label{feyn}
\end{center}
\end{figure}

In this case, we have elastic scattering of DM by nuclei through a
photon exchange (Fig.\ref{feyn}).  This is studied in detail for the
case of electric dipole moment in Section \ref{edm} and of magnetic
dipole moment in Section \ref{mdm}.

Next we consider the case of DM being  Majorana fermions.  These
have only non-zero transition moments between different mass
eigenstates.  Their interaction with photons is described by
\begin{equation}
\mathcal{L}_{inel}=-\frac{i}{2}\, \bar{\chi}_2
\sigma_{\mu\nu}(\mu_{12}+\gamma_5{ \mathcal D}_{12})\chi_1
\    \mathcal{F}^{\mu\nu}
\label{L_inelastic}
\end{equation}
where $\mu_{12}$ is the transition magnetic moment and ${ \mathcal D}_{12}$
is the transition electric moment.
 Majorana DM would have inelastic scattering off nuclei through a
photon exchange.  We study the case of transition electric dipole
moment in Section \ref{inel} and of transition magnetic dipole
moment in Section \ref{mdm}.

In Section \ref{relic} we calculate the bounds on the dipole moments
which would give the desired relic density abundance of DM.

In this paper we present two main types of results:

\begin{enumerate}
\item
We find bounds on dipolar moments, direct and transition, coming from WIMP search experiments, which update the results
in \cite{ksig04}. In addition we show that there are bounds on dipole moments from single photon search at LEP, an effect
not discussed in \cite{ksig04}.

\item
We find regions in parameter space that are consistent with the positive signal from DAMA
and with the null results from other experiments.

\end{enumerate}

All these results are presented in Section \ref{results}.  In
Appendix A we give some details of our DM scattering calculations,
and in Appendix B we show the DM annihilation cross section
calculations which are relevant for the relic density results.  In
Appendix C we review the contact scalar
interaction case, which is useful for comparison with our results.

\section{Electric Dipole Moment Interaction of Dark
Matter}\label{edm}

The differential cross section for a DM-proton elastic scattering via
interaction of the electric dipole moment of DM with the proton
charge  is given by
\begin{equation}\label{dRomega}
\frac{d\sigma}{dE_R}=\frac{e^2 \mathcal{D}^2}{4\pi v^2 E_R}
\end{equation}
where $\mathcal D$ is defined in (\ref{L_elastic}). Here $E_R$ is the
recoil energy of proton and $v$ is the speed of DM relative to the
nucleus.  The differential rate for nuclear scattering of DM through
electric dipole moment interaction for a nucleus of $Z$ protons and DM
of mass $m_\chi$ is given by
\begin{equation}\label{dR}
\frac{dR}{dE_R}=Z^2 N_T\,\frac{\rho_\chi}{m_\chi}
\int_{v>v_{min}}f(v)\,v\,\frac{d\sigma}{dE_R}\,dv
\end{equation}
where $N_T$ is the number of target nuclei in the detector and
$\rho_\chi$ is the local DM density.  The velocity distribution $f(v)$
and minimum speed of DM $v_{min}$, for a given energy threshold
$E_{Rmin}$ are given by \cite{gju96}
\begin{align}
f(v)&=\frac{4
v^2}{\sqrt{\pi}v_0^3}\,\exp\left(\frac{-v^2}{v_0^2}\right)\\
v_{min}&=\sqrt{\frac{m_N E_{Rmin}}{2 \mu_N^2}}
\end{align}
where $m_N$ is the mass of the target nucleus, $\mu_N=m_\chi
m_N/(m_\chi+m_N)$ is the reduced mass of the DM-nucleus system and
$v_0=220$ km/sec. The factor $Z^2$ appearing in eqn.(\ref{dR}) shows
the fact that this is a coherent scattering of the $Z$ protons in
the target nucleus.  To determine the expected event rate $R$ we
integrate eqn.(\ref{dR}) over the nuclear recoil energy:
\begin{equation}\label{R}
R=\int_{E_1/Q}^{E_2/Q}{dE_R\,\epsilon(QE_R)\frac{dR}{dE_R}}.
\end{equation}

\begin{figure}
\centering
\includegraphics[width=9cm,height=9cm]{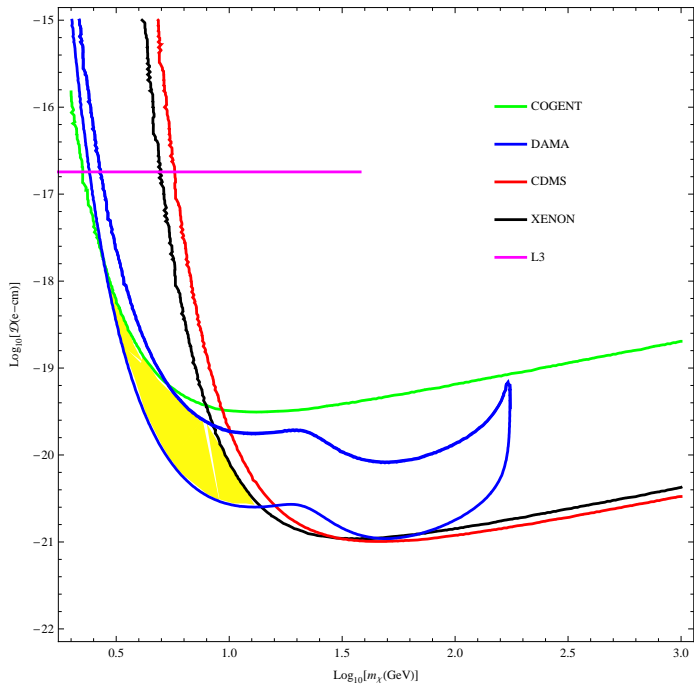}
\caption{Plot shows the allowed regions for DM electric dipole moment
with varying DM mass for elastic
scattering for different experiments.  Shaded region shows the allowed
parameter space for DAMA which
is consistent with all other experiments.}
\label{edm-plot}
\end{figure}

Here $\epsilon(QE_R)$ is the efficiency of the detector which depends
on the recoil energy and $Q$ is the quenching factor that relates the
observed energy with the actual recoil energy i.e $E_{Obs}=QE_R$.  $Q$
depends on the target nucleus and the nature of the detector.  $E_1$
and $E_2$ are limits of the observed energy interval in a direct
search experiment.  Therefore $E_1/Q$ and $E_2/Q$ give the
corresponding limits for the recoil energy $E_R$. Observed energies
are usually quoted in units of electron equivalent energies (keVee).
The detectors usually detect scintillation caused by electrons and
thereby measure energy transferred to electrons by the nucleus.  $Q$
then represents the efficiency with which the recoil energy of nucleus
is transferred to electrons which are detected by scintillation.  The
quenching factors for the different experiments analyzed here are
given in Table \ref{t}.

\begin{table}
\centering

\begin{tabular}{c c c}
\hline\hline\\
Experiment & Target nucleus & Quenching\\
&                & factor($Q$)\\
\hline\\
DAMA & Na & 0.3\\
& I  & 0.09\\
CDMS & Ge & 1\\
XENON& Xe & 1\\
COGENT&Ge & 0.2\\
\hline\hline
\end{tabular}
\caption{Quenching factors for different experiments used in this
analysis.}
\label{t}
\end{table}

In Fig.\ref{edm-plot} we show our results. The experiments give an
upper limit which is a function of the DM mass $m_\chi$, except for
DAMA whose result is represented as an allowed band. We would like
to point out that one can get a bound from the $e^+ e^- \rightarrow
\chi \chi \gamma$ process whose signal is a single photon detection
with missing energy. From the analysis of the collaboration L3
\cite{l3} we get $\mathcal{D}<6.6\times 10^{-16}$ e-cm, valid for
$m_\chi < 38$ GeV; this limit is also shown in Fig.\ref{edm-plot}.
We shall see the implications of this Figure in Section
\ref{results}; there we shall also discuss the issue of the DM relic
density.

\section{Inelastic Dark Matter and Electric Dipole Moment
Interaction}\label{inel}

In this Section we consider the case that the WIMP is a Majorana particle, with a
 a transition electric dipole moment $\mathcal{D}_{12}$ as defined
in (\ref{L_inelastic}).  In this case we can have inelastic
scattering $\chi_1 +N\rightarrow\chi_2+N$ where $\chi_1$ and $\chi_2$
are two different mass eigenstates, and  in general there is a
mass difference between $\chi_1$ and $\chi_2$, $\delta=m_2-m_1$.
Due to this mass difference, the minimum DM kinetic energy needed for the nucleon
scattering becomes higher. DAMA has lower detection threshold compared to most
experiments and therefore can be more sensitive to this scattering mode than the other
experiments. This was first proposed in \cite{weiner}.

\begin{figure}
\centering
\includegraphics[width=7.5cm,height=7.5cm]{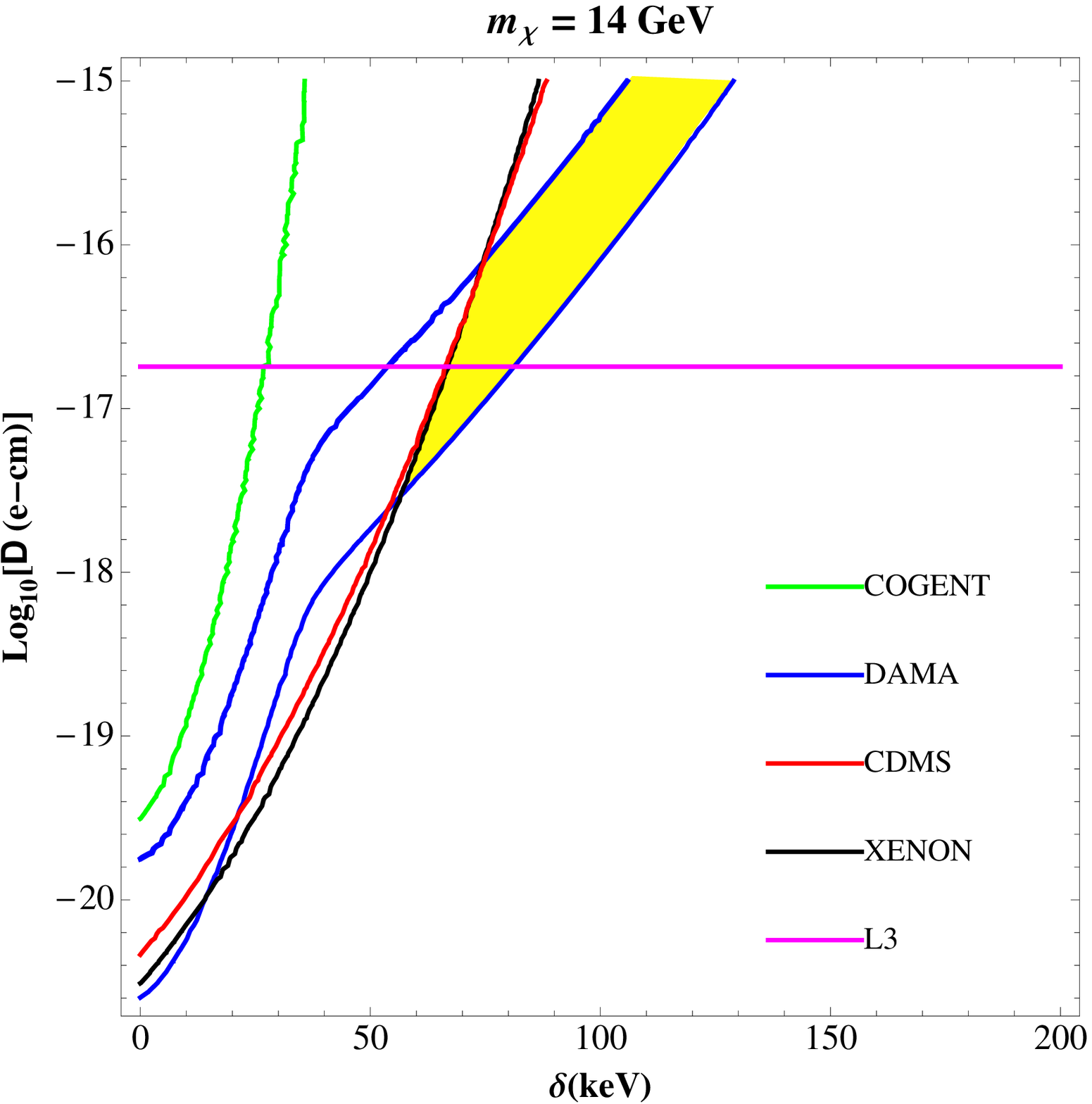}
\includegraphics[width=7.5cm,height=7.5cm]{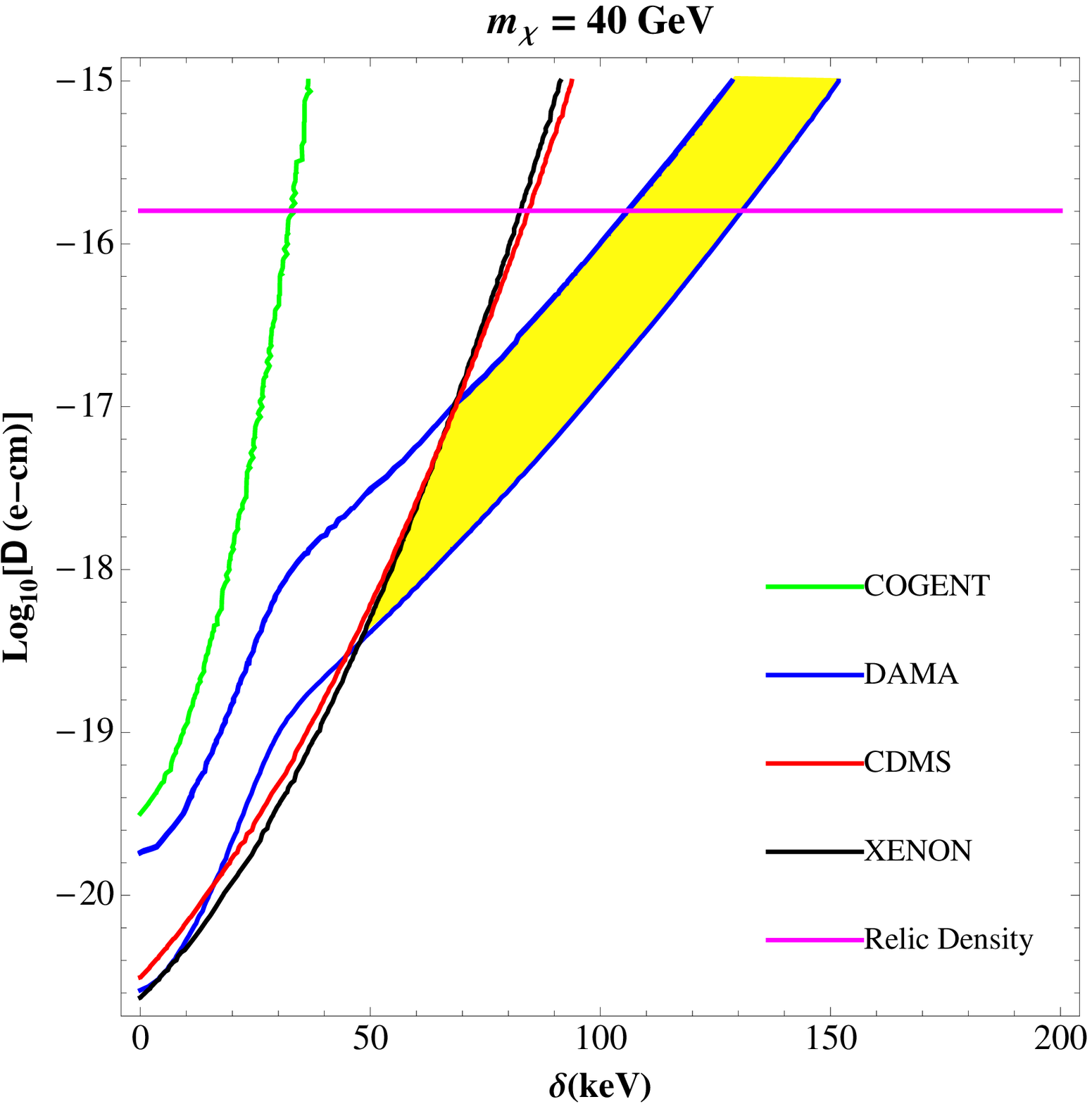}
\includegraphics[width=9cm,height=9cm]{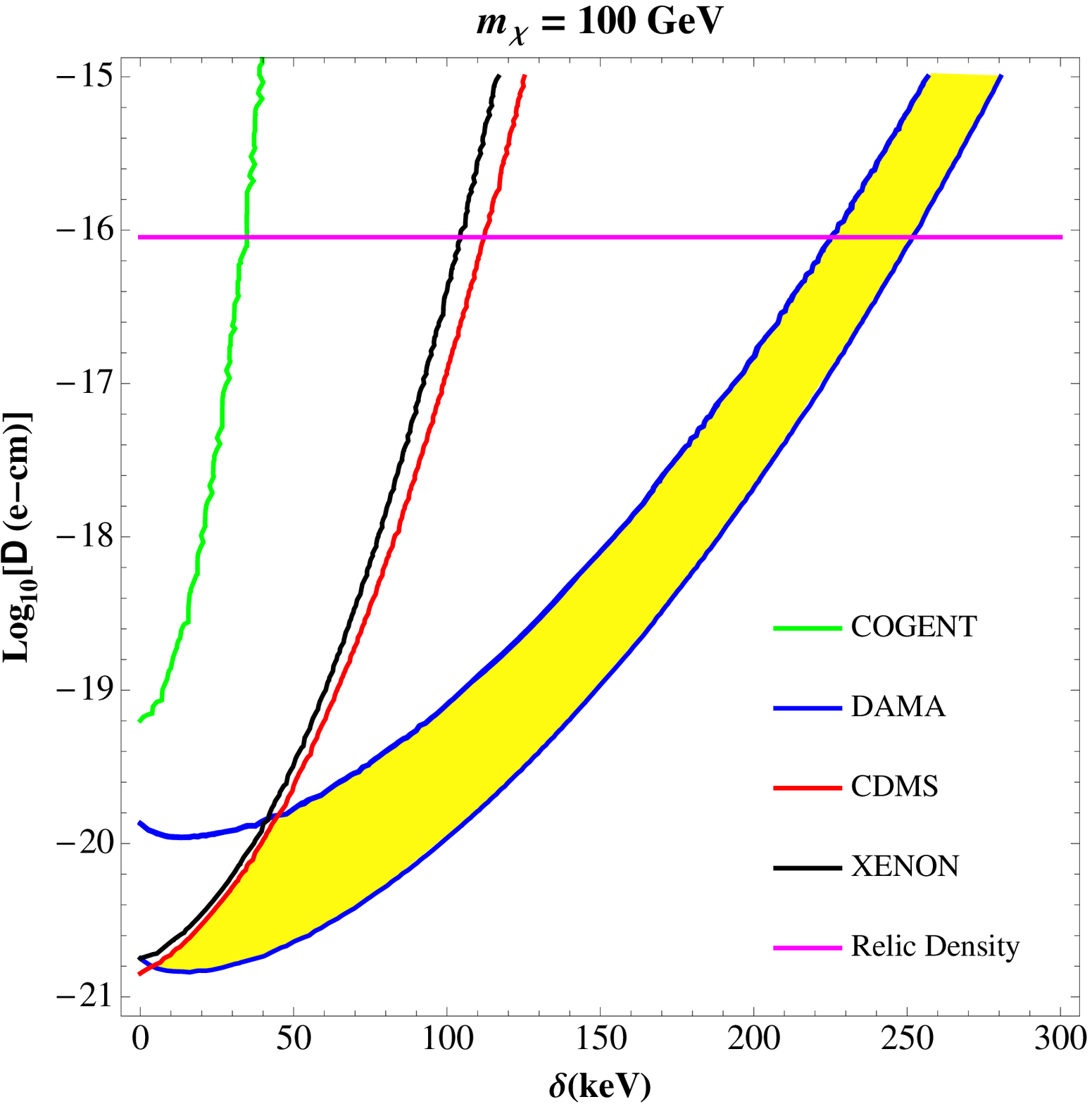}
\caption{Plot of DM electric dipole moment against the mass difference
$\delta$ for inelastic scattering for
different experiments.  Shaded region shows the allowed parameter
space for DAMA which is consistent with all other
experiments and in case of $m_\chi>38$ GeV it is also consistent
with the relic density.}
\label{edm-inel}
\end{figure}

\begin{figure}[t]
\centering
\includegraphics[width=9cm,height=9cm]{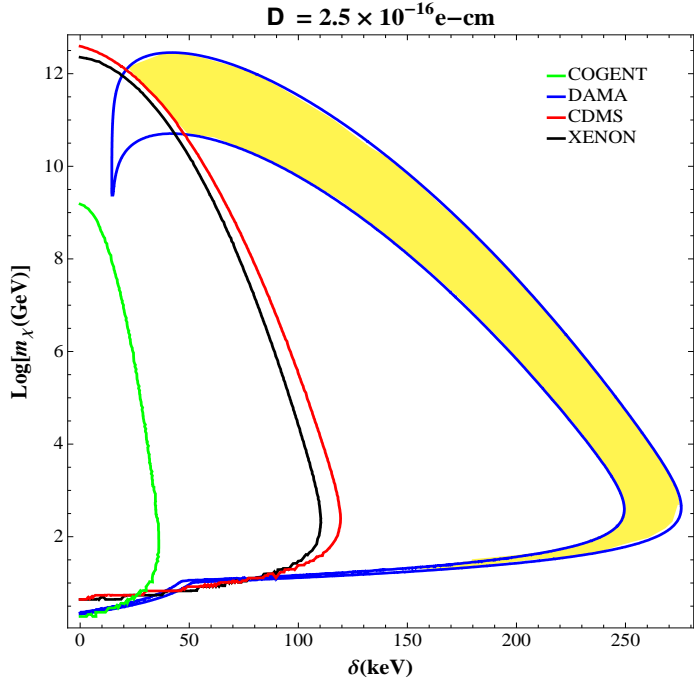}
\caption{Plot of allowed DM mass and mass splitting $\delta$ for a
fixed $\mathcal{D}$ which gives the correct relic density of DM.}
\label{edm-inel-fix}
\end{figure}

If we drop terms of higher order in $\delta$ (compared to $m_\chi$) we end up with
eqn.(\ref{dRomega}) for the differential cross section.  That is,
\begin{equation}\label{dcs}
 \frac{d\sigma_{inelastic}}{dE_R}=\frac{d\sigma_{elastic}}{dE_R}.
\end{equation}

However the total cross section in the elastic and inelastic case
are related to each other by a factor which enforces the condition
of the minimum kinetic energy required for inelastic scattering.
This relation has been derived at the end of Appendix A, and is given by,
\begin{equation}\label{el-inel}
 \sigma_{inelastic}=\sqrt{1-\frac{2\delta}{\mu_N v^2}}\sigma_{elastic}.
\end{equation}

But we do not use the total cross section in our analysis, only the
differential cross section and therefore the square root factor does not
figure in our calculations.  We integrate the differential cross section
per unit recoil energy over the energy interval in which the experimental data
have been observed.  The only change from the analysis in the elastic case is
that the minimum speed for scattering is now given by
\begin{equation}\label{vmin}
v_{min}=\sqrt{\frac{1}{2 m_N E_{Rmin}}}\left(\frac{m_N
E_{Rmin}}{\mu_N}+\delta\right)
\end{equation}
where notation has the same meaning as in the previous Section.

In Fig.\ref{edm-inel} we plot our results. We show a line for each
direct search experiment and the allowed range is on the right of
the line, except for DAMA that again is an allowed  band. In the
Figure we also plot the limit coming from LEP mentioned before,
$\mathcal{D}_{12}<6.6\times 10^{-16}$ e-cm, which is the same than
in the direct transition case if $|\delta| \ll m_\chi $. The
collider limit applies provided $m_\chi < 38$ GeV.

 Since in the inelastic case we have three parameters, namely,
 $\mathcal{D}_{12}, m_\chi$, and $\delta$, it is instructive to
 present Fig.\ref{edm-inel-fix} where we fix $\mathcal{D}_{12}$
 and show the constraint in the ($m_\chi, \delta$) plane. Here the
 allowed regions is on the right of the curves of the different
 experiments except for DAMA which is again an allowed band.

 We shall discuss the implications of all these limits in Section
 \ref{results} and especially for the relic density in Section
 \ref{relic}.

\section{Magnetic Dipole Moment Interaction}\label{mdm}

The differential cross section per unit energy transfer for elastic
scattering by a magnetic dipole moment interaction is given by
\begin{equation}
\frac{d\sigma}{dE_R}=\frac{e^2 \mu^2}{4\pi
E_R}\left(1+\frac{E_R}{2\mu_N v^2}\right).
\end{equation}

\begin{figure}[t]
\centering
\includegraphics[width=9cm,height=9cm]{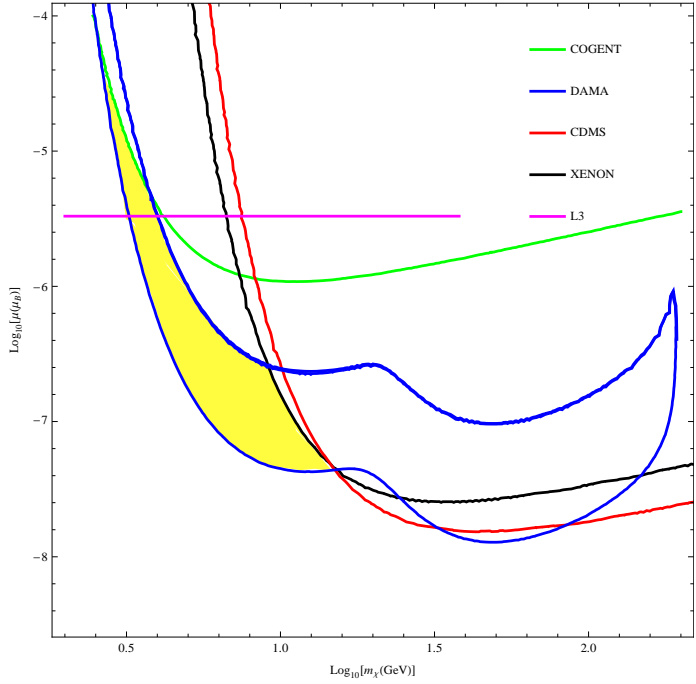}
\caption{Plot shows the allowed regions for DM magnetic dipole moment
with varying DM mass for elastic
scattering for different experiments.  Shaded region shows the allowed
parameter space for DAMA which
is consistent with all other experiments.}
\label{mu}
\end{figure}

\begin{figure}
\centering
\includegraphics[width=7.5cm,height=7.5cm]{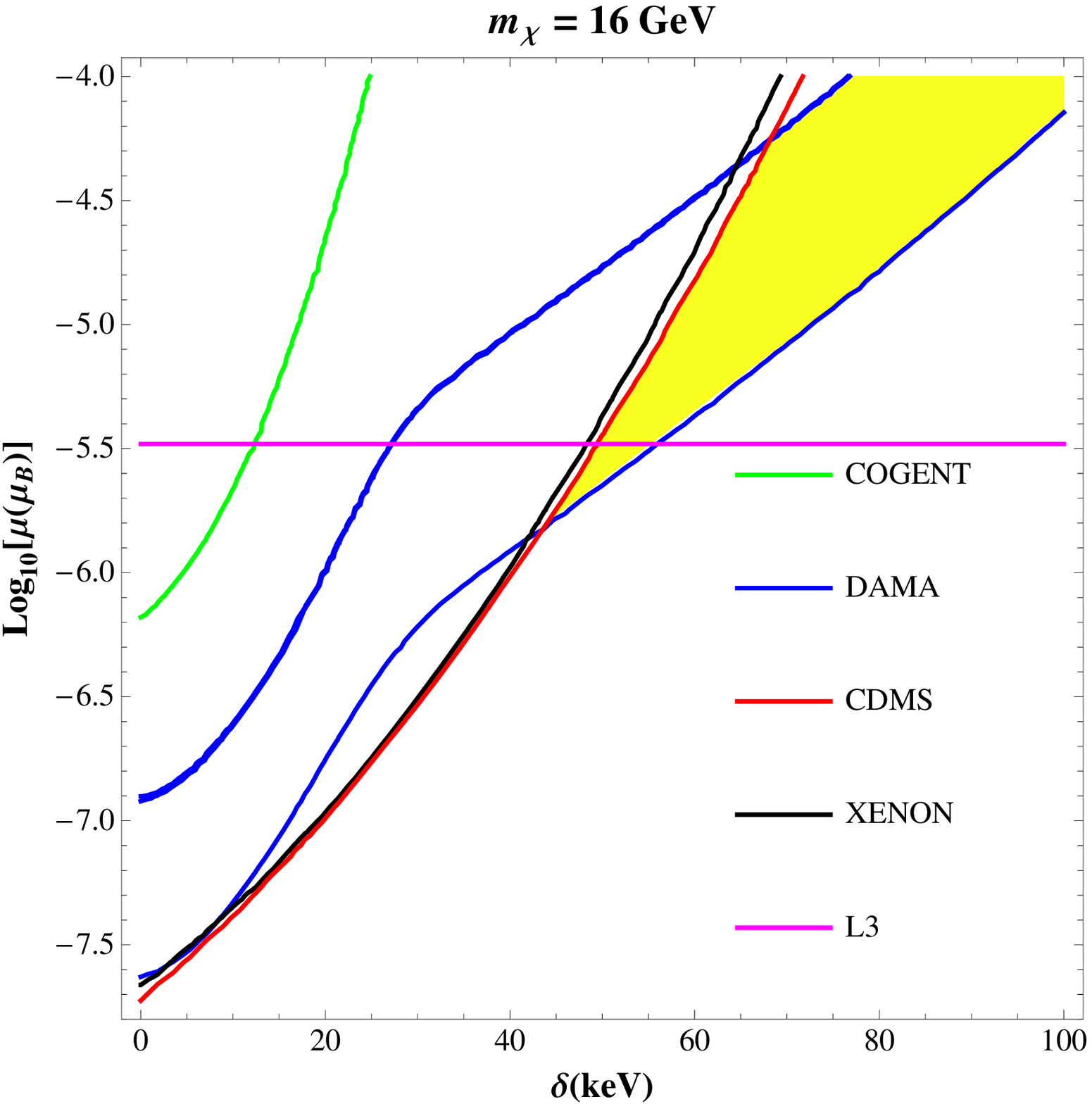}
\includegraphics[width=7.5cm,height=7.5cm]{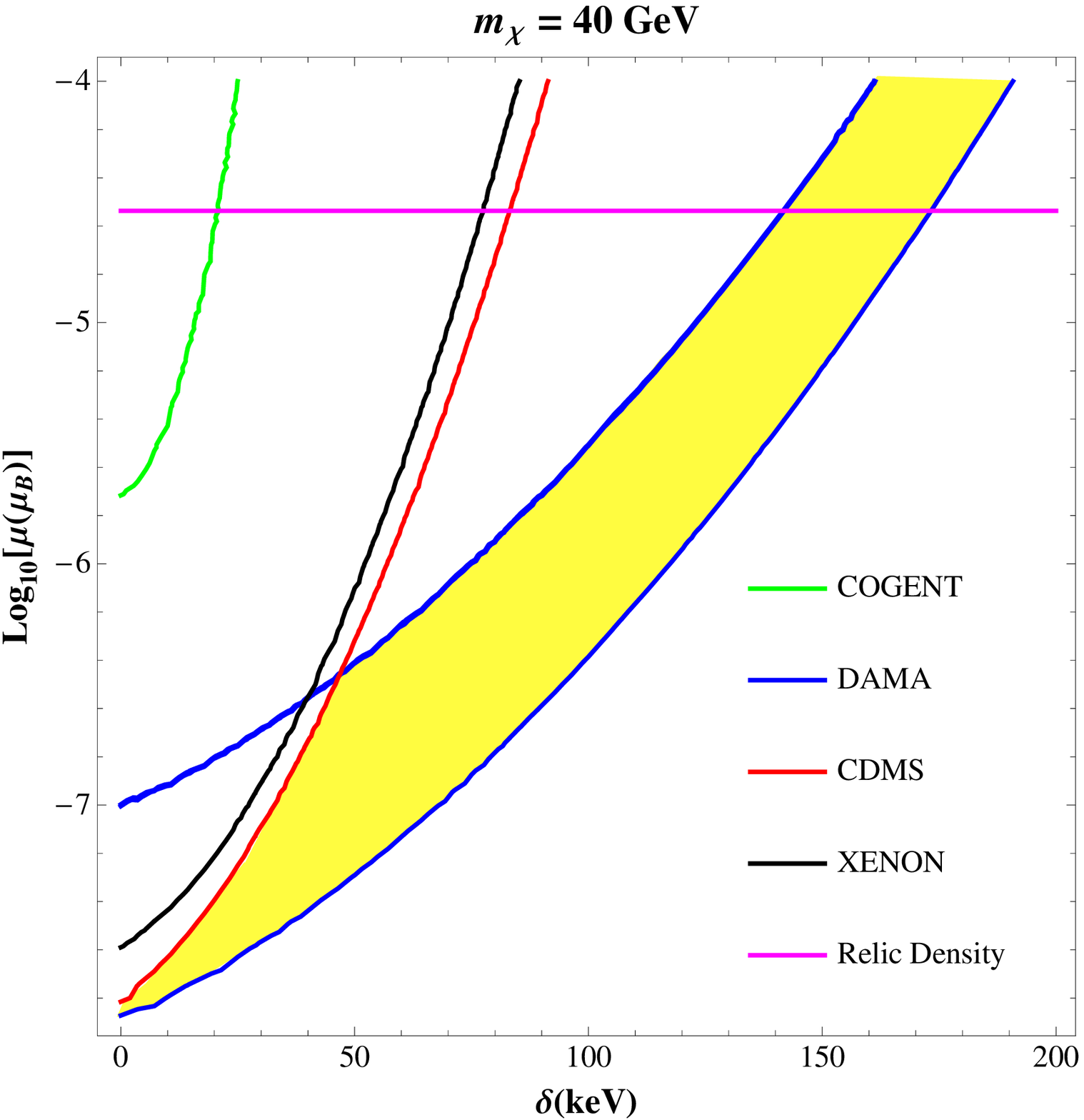}
\includegraphics[width=9cm,height=9cm]{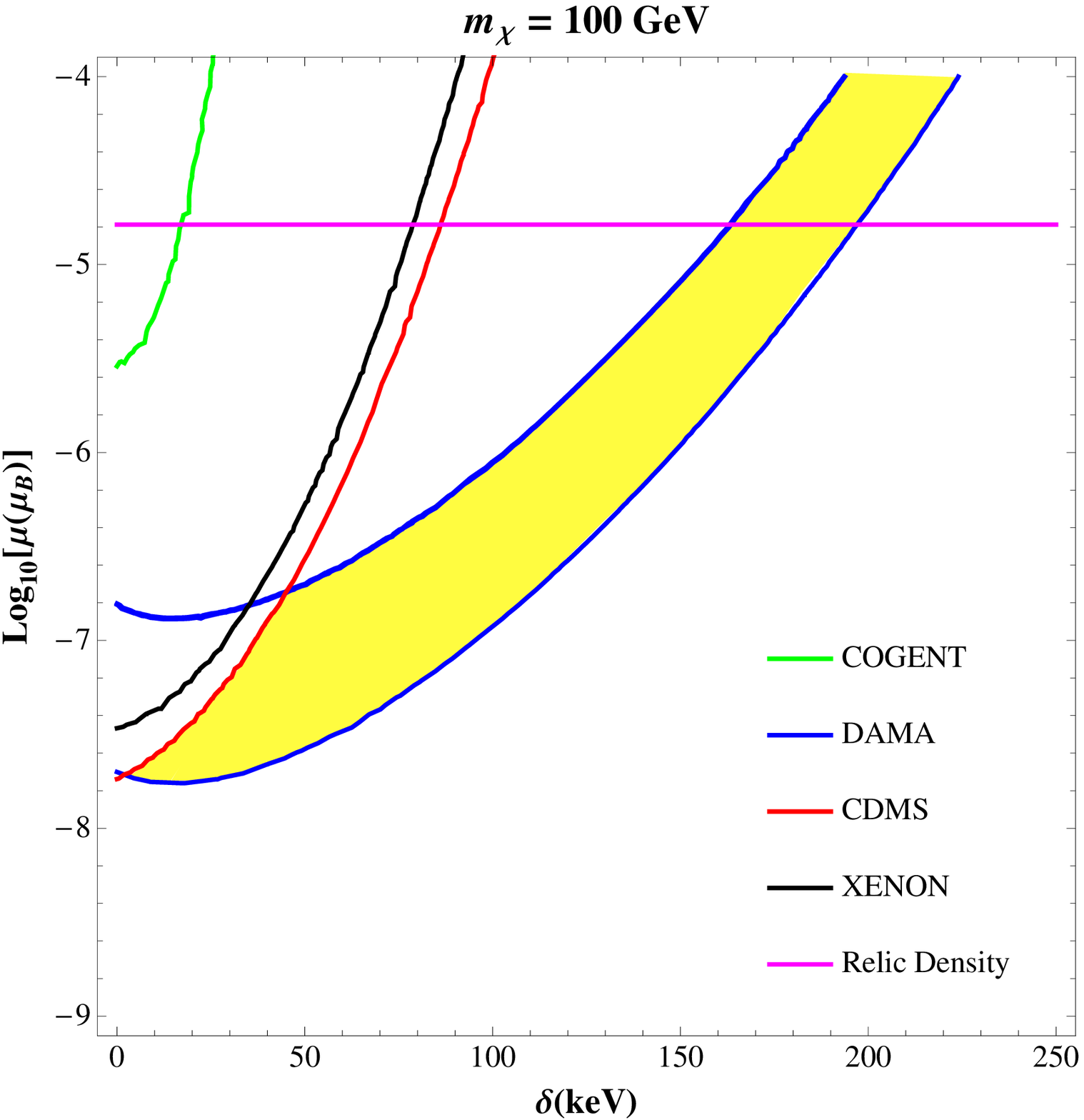}
\caption{Plot of DM magnetic dipole moment against the mass difference
$\delta$ for inelastic scattering for
different experiments.  Shaded region shows the allowed parameter
space for DAMA which is consistent with all other
experiments and in case of $m_\chi>38$ GeV it is also consistent
with the relic density.}
\label{mu-inel}
\end{figure}

\begin{figure}[t]
\centering
\includegraphics[width=9cm,height=9cm]{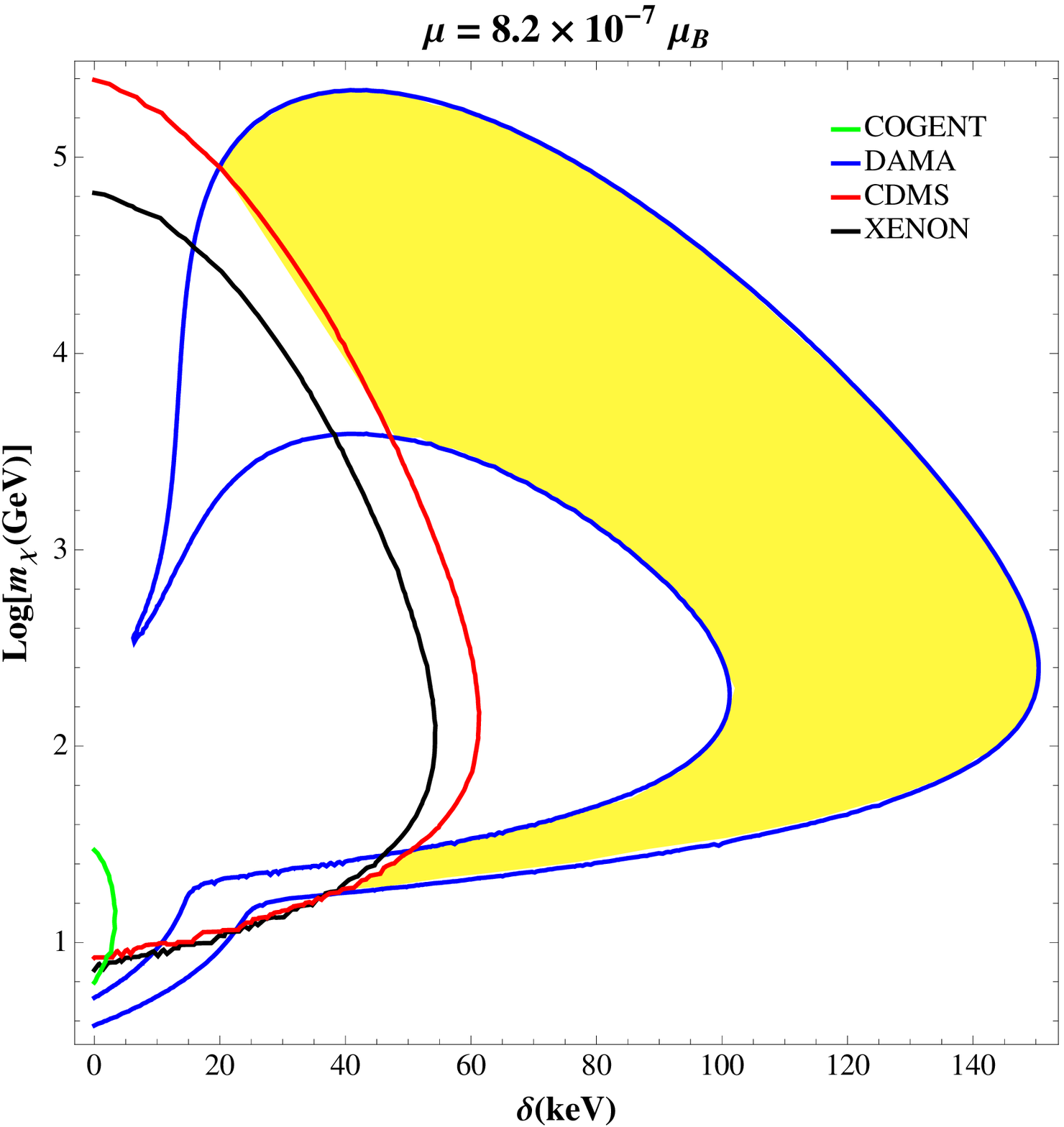}
\caption{Plot of allowed DM mass and mass splitting $\delta$ for a
fixed $\mu$ which gives the correct relic density of DM.}
\label{mu-inel-fix}
\end{figure}

In the case of inelastic scattering we get the same formula with $\mu$
substituted by the transition magnetic moment $\mu_{12}$, provided we
drop terms of higher order in the mass difference $\delta$.  Again,
the only
difference between elastic and inelastic is in the kinematics, because
the minimum speed for inelastic scattering is given by
eqn.(\ref{vmin}).  Fig.\ref{mu} and Fig.\ref{mu-inel} show the results
for magnetic dipole moment which are essentially the same as those for
electric dipole moment in terms of the allowed parameter space from
various experiments.
In Fig.\ref{mu-inel-fix} we plot the allowed regions in the space
($m_\chi$, $\mu_{12}$).

In the next section we discuss the implications of dipolar DM vis a
vis the DM relic density.


\section{Relic Density of Dipolar DM}\label{relic}

The relic abundance of dipolar DM is determined by the annihilation
cross section $\chi \bar \chi \rightarrow f \bar f$ through a dipole
vertex. In the case of Majorana fermions only non-identical fermions
can annihilate ($\chi_1 \bar \chi_2 \rightarrow f \bar f$) through
the dipole channel as only the transition dipole moments (${\cal
D}_{12}, \mu_{12}$) are non-zero. When the mass difference $\delta$
between $\chi_1$ and $\chi_2$ is small the cross section for
Majorana annihilation process is identical to that of the Dirac
fermions (with $ {\cal D}, \mu$ replaced with ${\cal D}_{12},
\mu_{12}$).

The annihilation cross section for electric dipole annihilation of
DM for $\chi \bar \chi \rightarrow f \bar f$ is given by (see appendix B)
\begin{equation}\label{ann-elec}
 \sigma v_{rel}=\frac{e^2\mathcal D^2}{48\pi} v_{rel}^2
\end{equation}
and the expression for the magnetic dipole annhilation case is
\begin{align}\notag
 \sigma v_{rel}&=\frac{e^2\mu^2}{4\pi}\left(1-\frac{v_{rel}^2}{6}\right)\\
               &\simeq\frac{e^2\mu^2}{4\pi}
 \label{ann-mag}
\end{align}
where $v=2\sqrt{1-\frac{4m_\chi^2}{s}}$ is the relative velocity of the two
annihilating WIMPs.  The thermal averaged cross section, $\langle \sigma
v_{rel} \rangle$, can be parameterized in terms of the temperature,
$T = m_\chi \langle v^2 \rangle /3$, where $v=v_{rel}/2$, such that
\begin{equation}
 \langle \sigma v_{rel} \rangle \equiv \sigma_0\left(\frac{T}{m_\chi}\right)^n
\end{equation}
where $n$ and the parameter $\sigma_0$ for magnetic and electric dipole
interactions take the following values
\begin{align}
 \mbox{Magnetic Dipole: }\quad\sigma_0&=\frac{e^2\mu^2}{4\pi}\,,\qquad n=0\\
 \mbox{Electric Dipole: }\quad\sigma_0&=\frac{3 e^2{\cal D}^2}{48\pi}\,,\qquad n=1
\end{align}

 Considering cold dark matter (CDM) we want to estimate the relic
density in the context of dipolar interactions of such CDM
candidates. The expression  for $x_f=\dfrac{m_\chi}{T_f}$, where the subscript
$f$ indicates the freeze out condition, is given up to a reasonable approximation
by \cite{kolb}
\begin{equation}
\begin{split}
 x_f=&\ln[0.038(n+1)(g/g_{*}^{1/2})m_{pl}m_\chi \sigma_0]\\
     &-\left(n+\frac{1}{2}\right)\ln\lbrace\ln[0.038(n+1)
       (g/g_{*}^{1/2})m_{pl}m_\chi \sigma_0]\rbrace
\end{split}
\end{equation}
where $m_{pl}$ is Planck mass, while $g$ and $g_{*s}$ are the effective number
of relativistic particles at the time of decoupling. The DM relic
density is then given by \cite{kolb}
\begin{equation}\label{omegah2}
 \Omega h^2=0.34\,\left(\frac{(n+1)x_f^{n+1}}{(g_{*s}/g_*^{1/2})}
                  \right)\,\frac{10^{-37}{\rm cm}^2}{\sigma_0}.
\end{equation}

The limit on cold dark matter density from WMAP \cite{wmap} is
$\Omega_{m} h^2=0.1099\pm 0.0062$.  Using (\ref{omegah2}), this gives
us the bound on ${\cal D}$, which gives the acceptable relic density
of dark matter, to be $\sim 2.5\times 10^{-16}$ e-cm. For the magnetic
dipole case again using (\ref{omegah2}) we find that the dipole
moment must be $\sim 8.2\times 10^{-7} \mu_B$.

For the case of non-identical Majorana annihilation the annihilation
cross sections are identical to the case of Dirac fermions and the
limits on the Dirac electric and magnetic moments from relic density
abundance are identical to the bounds on transition magnetic moments
given above.

\section{Results and Conclusions}\label{results}

\subsection{Direct dipole moments}

We have explored the parameter space of electric and magnetic dipole
moments of DM from the results of nuclear recoil experiments. We find
that the limit on direct electric dipole moment is $ {\cal D} < 1.6
\times 10^{-21}$ e-cm if the DM mass $m_\chi > 10$ GeV, see
Fig.\ref{edm-plot}. In this mass range there is no allowed parameter
space which is consistent with the positive signal from DAMA and with
the null results of other experiments.

For a DM mass $m_\chi =(3-13)
$ GeV and a moment ${\cal D} \sim 2.5\times 10^{-20}$ e-cm  there is a
region in parameter space where the DAMA signal is not ruled out by other
experiments, see Fig.\ref{edm-plot}.

\begin{figure}
\centering
\includegraphics[width=9cm,height=8cm]{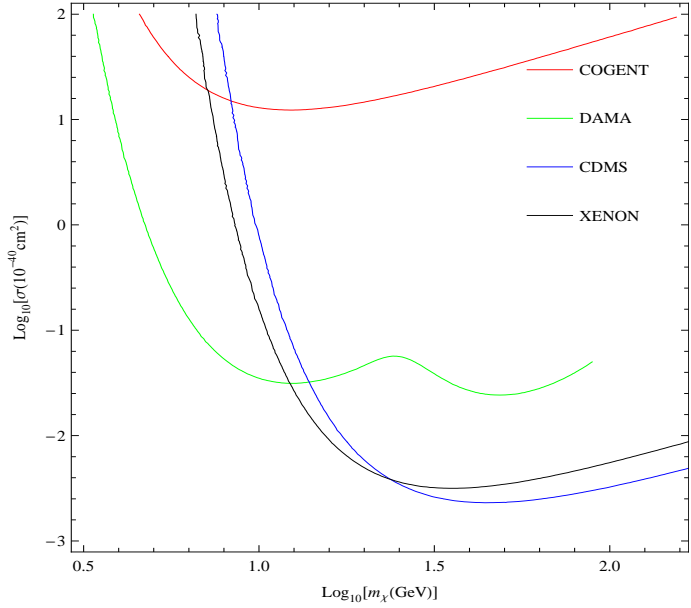}
\caption{Plot shows variation of spin-independent DM-nucleus
scattering cross section as a function of
the DM mass for different experiments.}
\label{sigma}
\end{figure}

For smaller masses, $m_\chi  < 3$ GeV, the Figure
shows that COGENT rules out DAMA.  In Appendix C we review the scalar interaction
mediated scattering (see Fig.\ref{sigma}) where we see that DAMA is
allowed by COGENT for any DM mass below about 15 GeV.  This can be attributed to the fact
that the DM interaction via electric dipole moment has a factor of
recoil energy ($E_R$ in the denominator of its cross section in
eqn.(\ref{dR}).  This means that for low threshold this cross section
is large and in the case of COGENT this threshold happens to be
smaller than that for DAMA.  But as the DM mass increases this
advantage in favour of COGENT is nullified by the DM mass factor which
again appears in the denominator of the cross section formulae.

The limit on direct magnetic dipole moment is $ \mu <  10^{-8} \mu_B$
if the DM mass $m_\chi > 25$ GeV, see Fig.\ref{mu}. In this mass range
there is no allowed parameter space which is consistent with a
positive signal from DAMA with null results of other experiments.
For DM masses $m_\chi =(3-12)$ GeV and $\mu \sim 6.3\times 10^{-7}
\mu_B$  there is a region in parameter space where the DAMA signal is not
ruled out by other experiments, see Fig.\ref{mu}. Actually, in the low DM mass range
the collider bound \cite{l3} is more restrictive than the direct WIMP search limits, as
can be see in Fig.\ref{mu}.

If the only interaction of DM with standard model particles had been
via electromagnetic dipole interaction then the relic density of the
DM particles would be determined uniquely from the annihilation
cross section of the EM-dipole mediated process $\chi \chi
\rightarrow f \bar f$ and $m_\chi$. Such a calculation shows
\cite{ksig04} that to get relic cold DM density consistent with
WMAP measurement the dipole moments have to be in the
range of $\mathcal{D} \simeq 5 \times 10^{-17} $ e-cm or $\mu \simeq
10^{-5} \mu_B$ when $m_\chi=(10-1000)$ GeV. The limits on DM dipole
moments that we obtain on the basis of nuclear scattering
experiments are much lower. This means that the DM particles must
have some other more dominant interaction which decouples at a
temperature $T_f \sim m_\chi/ 10$ to give the correct relic
abundance while the dipole interactions decouple much earlier and
have no bearing on the present DM density.  These other dominant
interactions however  must not dominate over dipole interactions in
nuclear scattering in order for our bounds on dipole moments to be
meaningful. A good example of such a situation is seen in certain
classes of WIMP models \cite{gju96}, where the interaction with
standard model particles can be by Higgs exchange which gives rise
to spin-independent (SI) interactions with nucleons or by $Z$ exchange
which gives rise to spin-dependent (SD) interactions \cite{barger}.
When the nuclear scattering is by SD interaction then there is no coherent
enhancement of the cross section by the atomic number. The bounds
on the cross section for SI interactions is therefore more
stringent than the bounds on SD interactions. The bounds for SI
interactions from nuclear scattering experiments is $\sigma_{SI} < 5
\times 10^{-44}$ cm$^2$  and for SD interactions it is $\sigma_{SD} <
5 \times 10^{-28}$ cm$^2$ (for $m_\chi \sim 100$ GeV). Our bounds are
relevant for WIMPs which can have a large spin-dependent cross
section $\sigma_{SD} \simeq 0.3\times 10^{-39}$ cm$^2$ to give the
correct relic density, but this cross section is too small to be
observed in nuclear scattering experiments.

\subsection{Transition dipole moments}

In Figs.\ref{edm-inel} and \ref{mu-inel}  we plot the constraints on the transition dipole moments ${\cal D}_{12}$
and ${\mu_{12}}$, respectively,
versus the DM mass $m_\chi$. We have fixed three values for the mass $m_\chi=14, 40$ and 100 GeV,
in order to illustrate our findings.

We see that a common feature of the plots is that the region where DAMA is consistent
with other experiments is quite large and extends to higher and higher inelasticities.
However for $m_\chi< 38$ GeV the region is further constrained by the collider bound from L3,
and the allowed region is reduced. The allowed area is more and more reduced as we lower $m_\chi$
and for $m_\chi < 10$ GeV there is no region at all. For $m_\chi> 38$ GeV there is no collider bound
and the region is large.

The most interesting aspect of the experimentally allowed parameter space for transition dipole moments is that the cosmologically preferred value (for getting the correct relic abundance) ${\cal D}_{ij} \sim 10^{-16}$ e-cm or $\mu_{ij}\sim 10^{-5} \mu_B$
\cite{ksig04} is consistent with the allowed values DAMA and other experiments when $m_\chi \gtrsim 38$ GeV and the mass split $\delta \simeq (50-100)$ keV.
Thus, we find that transition dipole scattering has a large parameter space where the
results of DAMA, the null results from other experiments and cosmological relic abundance
are all consistent.

The event rates depend upon $m_\chi $ and ${\cal D}$ or $\mu$ as ${\cal D}^2/m_\chi$ or $\mu^2/m_\chi$ respectively. The dipole moments can increase by four orders of magnitude from the ${\cal D}=10^{-20}$ e-cm to ${\cal D}=10^{-16}$ e-cm (and similarly for the case of magnetic moments) and the same event rates would be obtained  if mass is increased by eight orders of magnitude from $10^2$ GeV to $10^{10}$ GeV .  The range of DM mass for which  DAMA is  consistent with cosmology and L3 is therefore quite large, $38\,{\rm GeV} < m_\chi < 10^{10}\,{\rm Gev}$.
 This parameter space of $m_\chi$ and electric /magnetic moments can give rise to a missing energy signal via the process $p p\rightarrow \chi_1 \chi_2 \gamma +$hadronic jets, and the dipolar model of dark matter may be testable at the LHC.

\section*{Acknowledgements}
E.M. would like to thank Alessio Provenza for useful comments.
E.M. acknowledges support by the CICYT Research Project FPA
2008-01430 and the Departament d'Universitats, Recerca i Societat de
la Informaci\'{o} (DURSI), Project 2005SGR00916 and partly by the
European Union through the Marie Curie Research and Training Network
``UniverseNet'' (MRTN-CT-2006-035863).

\renewcommand{\theequation}{A\arabic{equation}}
\setcounter{equation}{0}
\section*{Appendix A: Cross Section Calculation for Electric and
Magnetic Dipole Moment Interactions with Nuclei}

Here we consider the elastic scattering process
$\chi\,+\,p\,\rightarrow\,\chi\,+\,p$.  We calculate the cross section
in the lab frame where the proton is at rest initially.  We assume
that even after the scattering the proton remains at rest since the
momentum transferred is very small in this process.  The entire
calculation is done in the non-relativistic limit.

\begin{center}
\textbf{Electric Dipole Moment}
\end{center}

The initial momenta of DM and proton are denoted by $k_i$ and $p_i$
respectively, while the final momenta are denoted by $k_f$ and $p_f$
respectively.  The amplitude squared for this process is given by
\begin{equation*}
\overline{|{\mathcal{M}}|^2}=-\frac{e^2\mathcal{D}^2 q_\alpha
q_\beta}{4q^4}
Tr[(\slashi{k_f}+m_\chi)\sigma^{\mu\alpha}\gamma_5
(\slashi{k_i}+m_\chi)\gamma_5\sigma^{\nu\beta}]\,\times
Tr[(\slashi{p_f}+m_p)\gamma_\mu(\slashi{p_i}+m_p)\gamma_\nu]
\end{equation*}
\begin{equation}\label{ampel}
\hskip-1.6cm=\frac{e^2\mathcal{D}^2}{4q^2}[64(q\cdot p_i)(k_i\cdot p_i)
+32m_\chi^2(q\cdot p_i)-16(q\cdot p_i)^2-8q^2(q\cdot p_i)-64(k_i\cdot p_i)^2]
\end{equation}
where $q$ is the momentum transferred, while $m_\chi$ and $m_p$ are
masses of DM and proton respectively.  In writing the above equation
we have made use of $k_f=k_i-q$ and $p_f=p_i+q$.
Here, $(q\cdot p_i)=m_p E_R$ and $k_i \cdot p_i \approx m_\chi m_p$.
Hence we can drop terms containing $(q\cdot p_i)$ and its higher orders,
keeping just the last term in eqn.(\ref{ampel}). This gives us
\begin{equation}
\overline{|{\mathcal{M}}|^2} =-\frac{16e^2\mathcal{D}^2}{q^2}(m_\chi
m_p)^2\,.
\end{equation}

Using $q^2=E_R^2-2 m_p E_R\approx -2m_p E_R$ we finally arrive at the
following expression for the amplitude squared,
\begin{equation}\label{ampel1}
\overline{|{\mathcal{M}}|^2}=\frac{8e^2\mathcal{D}^2}{E_R}m_\chi^2 m_p\,.
\end{equation}

Now, the differential cross section is given by
\begin{align}\label{ds}\notag
 d\sigma&=\frac{1}{2m_p 2E_1 v}\,\frac{d^3p_f}{(2\pi)^3E_p}
          \,\frac{d^3k_f}{(2\pi)^3E_2}\,(2\pi)^4\,
          \delta^4(p_i+k_i-p_f-k_f)\,
         \overline{|{\mathcal{M}}|^2}\\\notag
        &=\frac{1}{64\pi^2 m_p E_1 v}\,\frac{d^3 k_f}{E_2 E_p}
          \,\delta(m_p+E_1-E_p-E_2)\,
         \overline{|{\mathcal{M}}|^2}\\
        &=\frac{1}{64\pi^2 m_p E_1 v}\,\frac{|\vec{k_f}|}{E_p}\,d\Omega
          \,\overline{|{\mathcal{M}}|^2}
\end{align}
where $E_1$ and $E_2$ are energies of the initial and final DM states.
Also the recoil energy can be written as follows:
\begin{equation*}
 E_R=\frac{|\vec{q}|^2}{2m_p}=\frac{|\vec{k_i}|^2+|\vec{k_f}|^2
                              -2|\vec{k_i}||\vec{k_f}|\cos\theta}{2 m_p}
\end{equation*}
therefore,
\begin{equation*}
 dE_R=-\frac{|\vec{k_i}||\vec{k_f}|}{m_p}d(\cos\theta).
\end{equation*}

And we know that $d\Omega=-2\pi d(\cos\theta)$, so we can write
\begin{equation}
 d\Omega=2\pi\frac{m_p}{|\vec{k_i}||\vec{k_f}|}dE_R,
\end{equation}

Using this in eqn.(\ref{ds}) we get
\begin{align}\notag
 \frac{d\sigma}{dE_R}&=\frac{1}{32\pi E_1 v}\,\frac{1}{E_p|\vec{k_i}|}
                      \,\overline{|{\mathcal{M}}|^2}\\
                     &=\frac{1}{32\pi E_1^2 E_p v^2}\,
                     \overline{|{\mathcal{M}}|^2}.
\end{align}

We use the approximations, $E_1^2\approx m_\chi^2$ and $E_p\approx m_p$,
which gives us
\begin{equation}\label{ds1}
 \frac{d\sigma}{dE_R}=\frac{1}{32\pi m_\chi^2 m_p v^2}\,\overline{|{\mathcal{M}}|^2}.
\end{equation}

Substituting for $\overline{|\mathcal{M}|^2}$ from eqn.(\ref{ampel1}) we get
\begin{equation}
  \frac{d\sigma}{dE_R}=\frac{e^2\mathcal{D}^2}{4\pi E_R v^2}
\end{equation}
which is same as eqn.(\ref{dRomega}).

\begin{center}
\textbf{Magnetic Dipole Moment }
\end{center}

The trace in this case is same as in the case of EDM but without any
$\gamma_5$ in it.  The amplitude squared for this process is given by
\begin{equation*}
\hskip-.6cm
\overline{|{\mathcal{M}}|^2}=\frac{e^2\mu^2 q_\alpha
q_\beta}{4q^4}
Tr[(\slashi{k_f}+m_\chi)\sigma^{\mu\alpha}
(\slashi{k_i}+m_\chi)\sigma^{\nu\beta}]\,\times
Tr[(\slashi{p_f}+m_p)\gamma_\mu(\slashi{p_i}+m_p)\gamma_\nu]
\end{equation*}
\begin{equation}
\begin{split}
&=\frac{4 e^2 \mu^2}{q^2} \left[4(q\cdot p_i)(k_i\cdot p_i) - 4(q\cdot
p_i)m_\chi^2
- (q\cdot p_i)^2 - 4(k_i\cdot p_i)^2 + 4 m_\chi^2 m_p^2\right]\\
&- \frac{4 e^2 \mu^2}{q^4} \left[4(q\cdot p_i)^2 m_\chi^2\right].
\end{split}
\end{equation}

Here we drop terms which are higher order in $q\cdot p_i$ except the
last term which is enhanced by a factor of $q^2$ in the denominator.
Thus we get
\begin{equation}\label{m2}
\overline{|\mathcal{M}|^2} = \frac{16 e^2 \mu^2}{q^2}
\left[(q\cdot p_i)(k_i\cdot p_i)-(q\cdot p_i)m_\chi^2
-(k_i\cdot p_i)^2 + m_\chi^2 m_p^2-\frac{(q\cdot p_i)^2m_\chi^2}{q^2}\right]
\end{equation}
where
\begin{align*}
q\cdot p_i&=m_p E_R\\
k_i\cdot p_i &\approx m_\chi m_p\\
(k_i\cdot p_i)^2 &\approx m_\chi^2 m_p^2+m_\chi^2 m_p^2 v^2
\end{align*}
Using these results in eqn.(\ref{m2}) we get
\begin{equation}
\overline{|{\mathcal{M}}|^2}=\frac{16 e^2 \mu^2}{q^2}\left[m_\chi m_p^2 E_R
                             -m_p m_\chi^2 E_R - m_\chi^2 m_p^2 v^2
                             -\frac{m_\chi^2 m_p^2 E_R^2}{q^2}\right].
\end{equation}

Again using $q^2\approx -2m_p E_R$ gives
\begin{align}\notag
\overline{|{\mathcal{M}}|^2}&=\frac{8 e^2 \mu^2}{E_R} \left[m_\chi^2 m_p v^2 + m_\chi^2 E_R
                             - m_\chi m_p E_R - \frac{m_\chi^2 E_R}{2} \right]\\
&=\frac{8 e^2 \mu^2}{E_R} \left[m_\chi^2 m_p v^2 + \frac{m_\chi^2 E_R}{2}
                             - m_\chi m_p E_R \right].
\end{align}

The differential cross section per unit recoil energy is given
by eqn.(\ref{ds1}).  Substituting from the above equation in eqn.(\ref{ds1})
we get
\begin{align*}
\frac{d\sigma}{dE_R}&=\frac{e^2\mu^2}{4\pi E_R}
                      \left[1+\frac{E_R}{2m_p v^2}-\frac{E_R}{m_\chi v^2}\right]\\
                    &=\frac{e^2\mu^2}{4\pi E_R}
                      \left[1+\frac{E_R}{2\mu_p v^2}-\frac{3E_R}{2m_\chi v^2}\right]
\end{align*}
where $\mu_p=\dfrac{m_\chi m_p}{m_\chi+m_p}$ is the reduced mass of the DM-proton
system.  The last term in the above equation can be dropped for large values of DM
mass.  And so we get the following expression for the differential cross section
per unit energy,
\begin{equation}
 \frac{d\sigma}{dE_R}=\frac{e^2\mu^2}{4\pi E_R}
                      \left[1+\frac{E_R}{2\mu_p v^2}\right].
\end{equation}

\begin{center}
 {\bf Total Cross Section for Inelastic Scattering}
\end{center}

In the case of inelastic scattering we consider the WIMPs as Majorana
particles.  As mentioned before in the main section that the differential
scattering cross section in both elastic and inelastic scattering
is the same.  But the total cross sections in the two cases are related to
each other as in eqn.(\ref{el-inel}).  Now for inelastic scattering we have,
\begin{align*}
 k_1^2-k_2^2&=2\mu_N\delta\\
 \mbox{or }k_2^2&=k_1^2-2\mu_N\delta.
\end{align*}

The magnitude of the momentum transferred (working in CM frame) is then,
\begin{align*}\label{magq}
 q^2&=k_1^2+k_2^2-2k_2k_2\cos\theta\\
 &=2k_1^2\left(1-\sqrt{1-\frac{2\delta}{\mu_N v^2}}\cos\theta\right)-2\mu_N\delta
\end{align*}
where we have used $k_1^2=\mu_N^2 v^2$.  And so,
\begin{equation*}
 dq^2=-2k_1^2\sqrt{1-\frac{2\delta}{\mu_N v^2}}d(\cos\theta).
\end{equation*}

But $dq^2=2m_N dE_R$, therefore
\begin{equation*}
 dE_R=-\frac{k_1^2}{m_N}\sqrt{1-\frac{2\delta}{\mu_N v^2}}d(\cos\theta).
\end{equation*}

The above equation in the elastic case becomes
\begin{equation*}
 dE_R=-\frac{k_1^2}{m_N}d(\cos\theta).
\end{equation*}

Now the total cross section for elastic scattering is given by,
\begin{align*}
 \sigma_{elastic}&=\int\frac{d\sigma_{elastic}}{dE_R}dE_R\\
 &=\frac{2\pi k_1^2}{m_N}\int\frac{d\sigma_{elastic}}{d\Omega}d(\cos\theta).
\end{align*}

For the inelastic case we have,
\begin{align*}
 \sigma_{inelastic}&=\frac{2\pi k_1^2}{m_N}\sqrt{1-\frac{2\delta}{\mu_N v^2}}
                     \int\frac{d\sigma_{elastic}}{d\Omega}d(\cos\theta)\\
                   &=\sqrt{1-\frac{2\delta}{\mu_N v^2}}\sigma_{elastic}.
\end{align*}


\renewcommand{\theequation}{B\arabic{equation}}
\setcounter{equation}{0}
\section*{Appendix B: Annihilation Cross Section of Majorana DM}

The process involved here is $\chi_1(k_1)\bar{\chi}_2(k_2)\rightarrow f(p_1)\bar{f}(p_2)$,
$f$ being a fermion of charge $e$ with negligible mass.
Now we work in the CM frame where,
\begin{align*}
 k_1&=(E_1,0,0,k), &&k_2=(E_2,0,0,-k),\\
 p_1&=(E_f,E_f\sin\theta,0,E_f\cos\theta), &&p_1=(E_f,E_f\sin\theta,0,E_f\cos\theta),\\
 q&=p_1+p_2, &&q^2=4E_f^2.
\end{align*}

The amplitude squared for this process is given by
\begin{equation}\label{m2i}
 |\mathcal M|^2=\frac{e^2g^2}{q^4}q^\alpha q^\beta
                 [v(k_2)\bar v(k_2)\Gamma_{\mu\alpha}
                  u(k_1)\bar u(k_1)\Gamma_{\nu\beta}]
                 [u(p_1)\bar u(p_1)\gamma^\mu
                  v(p_2)\bar v(p_2)\gamma^\nu]
\end{equation}
where $\Gamma_{\mu\alpha}=\sigma_{\mu\alpha}$ and $g=\mu$ for
magnetic dipole interaction, whereas for electric dipole interaction
$\Gamma_{\mu\alpha}=\sigma_{\mu\alpha}\gamma_5$ and $g=\mathcal D$.
For Majorana particles,
\begin{equation}
\label{majorana}
 \psi(x)=\sum_{s=1,2}\int \frac{d^3 k}{(2\pi)^3 2k_0}[b_s(k)u_s(k)e^{-ikx}+b_s^\dagger v_s(k)e^{ikx}]\\
\end{equation}

In the Majorana representation of $\gamma$-matrices, $C=-\gamma_0$
and the Majorana condition implies that $\psi=\psi^c=C \bar
\psi^T=\psi^\ast$. Imposing $\psi=\psi^\ast$ on (\ref{majorana}) we
get  the relation $u_s=v_s^*$. Using this relation we can rewrite
eqn.(\ref{m2i}) as
\begin{align}\label{m2maj}
 |\mathcal M|^2&=\frac{e^2g^2}{q^4}q^\alpha q^\beta
                 [v(k_2)u^T(k_2)\gamma_0\Gamma_{\mu\alpha}
                  u(k_1)v^T(k_1)\gamma_0\Gamma_{\nu\beta}]
                 [u(p_1)\bar u(p_1)\gamma^\mu
                  v(p_2)\bar v(p_2)\gamma^\nu].
\end{align}

Now the spin sum rules for Majorana fermions are given by \cite{haber}
\begin{align}\label{srule1}
 \sum_s u_{s}(p)v^T_s(p)&=(\slashi p+m)C^T\\\label{srule2}
 \sum_s v_{s}(p)u^T_s(p)&=(\slashi p -m)C^T
\end{align}
where $C^T=\gamma_0$.
We then take the sum over spins of eqn.(\ref{m2maj}) and use eqns.(\ref{srule1}) and
(\ref{srule2}) to obtain
\begin{equation}\label{majav}
 \overline{|\mathcal M|^2}=\frac{e^2g^2}{4q^4}q^\alpha q^\beta\,
                 Tr[(\slashi{k_2}-m_2)\Gamma_{\mu\alpha}(\slashi{k_1}+m_1)\Gamma_{\nu\beta}]\,
                 Tr[\slashi{p_1}\gamma^\mu\slashi{p_2}\gamma^\nu]
\end{equation}
where $m_1$ and $m_2$ are masses of $\chi_1$ and $\chi_2$ respectively.
Now we discuss the two cases, beginning with the electric dipole case first.
\begin{center}
 \textbf{Electric Dipole interaction}
\end{center}
Eqn.(\ref{majav}) with $g={\cal D}$ and $\Gamma_{\mu \nu}=
\sigma_{\mu \nu} \gamma_5$ yields the following expression for the
averaged amplitude squared,
\begin{equation}\label{mjed}
 \overline{|\mathcal M|^2}=\frac{e^2\mathcal D^2}{4q^4}(p_1\cdot p_2)
                           [(k_1\cdot p_1)(k_1\cdot p_2)+(k_2\cdot p_1)(k_2\cdot p_1)
                            -(p_1\cdot p_2)m_1 m_2]
\end{equation}
where $q=p_1+p_2$.  Now in CM frame,
\begin{align*}
 k_1\cdot p_1&=E_1E_f-kE_f\cos\theta&&k_1\cdot p_2=E_1E_f+kE_f\cos\theta\\
 k_2\cdot p_1&=E_2E_f+kE_f\cos\theta&&k_2\cdot p_2=E_2E_f-kE_f\cos\theta\\
 p_1\cdot p_2&=\frac{q^2}{2}&&q^2=4E_f^2
\end{align*}
Using these in eqn.(\ref{mjed}) we get
\begin{align}\notag
 \overline{|\mathcal M|^2}&=4e^2\mathcal D^2[E_1 E_2-k^2\cos^2\theta-m_1 m_2]\\\label{mjed2}
                          &=4e^2\mathcal D^2 E_1 E_2\left[\left(1-\frac{m_1 m_2}{E_1 E_2}\right)
                            +\left(1-\frac{(s-m_1^2-m_1^2)}{2E_1 E_2}\right)\cos^2\theta\right]
\end{align}
where we use $k=\dfrac{s}{2}-E_1 E_2-\dfrac{(m_1^2+m_2^2)}{2}$ with $s=(E_1+E_2)^2$.

The differential cross section in the CM frame is given by
\begin{align}\notag
 \frac{d\sigma}{d\Omega}&=\frac{1}{64\pi^2 E_1 E_2 v_{rel}}\frac{E_f}{\sqrt s}
                         \overline{|\mathcal M|^2}\\
 &=\frac{e^2\mathcal D^2}{32\pi^2 v_{rel}}\left[\left(1-\frac{m_1 m_2}{E_1 E_2}\right)
   +\left(1-\frac{(s-m_1^2-m_2^2)}{2E_1 E_2}\right)\cos^2\theta\right]
\end{align}
where $E_f=\dfrac{\sqrt s}{2}$.  Integrating above equation over $d\Omega$ gives
the total cross section as
\begin{equation}
 \sigma v_{rel}=\frac{e^2\mathcal D^2}{8\pi}\left[\left(1-\frac{m_1 m_2}{E_1 E_2}\right)
 +\frac{1}{3}\left(1-\frac{(s-m_1^2-m_2^2)}{2E_1 E_2}\right)\right].
\end{equation}

Eqn.(\ref{ann-elec}) can be obtained by taking $m_1=m_2$ and $E_1=E_2$
in the above equation.

\begin{center}
 \bf{Magnetic Dipole Interaction}
\end{center}

For the magnetic dipole case,  with $g={\mu}$ and $\Gamma_{\mu \nu}=
\sigma_{\mu \nu}$ in eqn.(\ref{majav}), the averaged amplitude squared reads
\begin{equation}
 \overline{|\mathcal M|^2}=\frac{e^2\mu^2}{4q^4}(p_1\cdot p_2)
                           [(k_1\cdot p_1)(k_1\cdot p_2)+(k_2\cdot p_1)(k_2\cdot p_1)
                            +(p_1\cdot p_2)m_1 m_2]
\end{equation}
and this equation in the CM frame looks like

\begin{equation}
 \overline{|\mathcal M|^2}=4e^2\mu^2 E_1 E_2\left[\left(1+\frac{m_1 m_2}{E_1 E_2}\right)
                           +\left(1-\frac{(s-m_1^2-m_1^2)}{2E_1 E_2}\right)\cos^2\theta\right].
\end{equation}

The total cross section is then given by
\begin{equation}
 \sigma v_{rel}=\frac{e^2\mu^2}{8\pi}\left[\left(1+\frac{m_1 m_2}{E_1 E_2}\right)
 +\frac{1}{3}\left(1-\frac{(s-m_1^2-m_2^2)}{2E_1 E_2}\right)\right].
\end{equation}

One can obtain eqn.(\ref{ann-mag}) by taking $m_1=m_2$ and $E_1=E_2$
in the above formula.

\renewcommand{\theequation}{C\arabic{equation}}
\setcounter{equation}{0}
\section*{Appendix C: Event Rate for DM-Nucleus Elastic Scattering
through Scalar Interaction}

Here we consider scattering of DM and nucleus through scalar mediated
spin-independent interaction.  The details of this calculation are
given in \cite{gju96}.  Here we shall give only the results of the
calculation.  The differential recoil rate per unit detector mass for
DM-proton scattering can be written as
\begin{equation}\label{dsig}
\frac{dR}{dE_R}=\frac{\rho_\chi\sigma_p
A^2}{2m_\chi\mu_p^2}\int_{v>v_{min}}{dv\frac{f(v)}{v}}
\end{equation}
where $q=\sqrt{m_N E_R}$ is the nucleus recoil momentum, $\sigma_p$ is
the DM-proton cross-section, $A$ is the number of nucleons and $\mu_p$
is the reduced mass of the DM-proton system.  Here $F(E_R)$ is the
Woods-Saxon form factor given by
\begin{equation}
F(E_R)=\frac{3j_1(qr)}{qr}e^{-q^2 s^2}
\end{equation}
where $q=\sqrt{2m_N E_R}$ is the momentum transferred, $m_N$ being the
nucleus mass, $s\simeq 1 $ fm and $r=\sqrt{R^2-5s^2}$ with $R=1.2
A^{1/3}$ fm.  The spin-independent differential cross section for
contact interaction scattering is given by
\begin{align}\label{dsig1}
\frac{d\sigma}{d|\mathbf{q}|^2}&=G_F^2 \frac{C}{v^2}F^2(E_R)
=\frac{\sigma_0}{4\mu_N^2 v^2}F^2(E_R)\\\notag
\mbox{where}&\\
&C=\frac{1}{\pi G_F^2}\left[Zf_p+(A-Z)f_n\right]^2\\\label{sig0}
&\sigma_0=\int_0^{4\mu_N^2
v^2}\quad{\frac{d\sigma(q=0)}{d|\mathbf{q}|^2}d|\mathbf{q}|^2}
=\frac{4\mu_N^2}{\pi}\left[Zf_p+(A-Z)f_n\right]^2
\end{align}

Here $|\mathbf{q}|^2=2m_N E_R$ is the magnitude square of the momentum
tranferred, $\mu_N$ is the reduced mass of the DM-nucleus system, $Z$
is the number of protons and $A$ is the number of nucleons.  Taking
$f_p\sim f_n$ we can write the DM-nucleus cross section in terms of
the DM-proton cross section\cite{gelmini} using eqn.(\ref{sig0}) as
\begin{equation}\label{sigp}
\sigma_0=\frac{\mu_N^2}{\mu_p^2}A^2\sigma_p
\end{equation}
where $\mu_p$ is the reduced mass of the proton-DM system.
 Substituting from eqn.(\ref{sigp}) in eqn.(\ref{dsig1}) we
have
\begin{equation}\label{dsig2}
\frac{d\sigma}{d|\mathbf{q}|^2}=\frac{\sigma_p A^2}{4\mu_p^2
v^2}F^2(E_R)
\end{equation}

The differential rate per unit detector mass can be written as
\begin{equation}
dR=\frac{\rho_\chi}{m_\chi
m_N}\left(\frac{d\sigma}{d|\mathbf{q}|^2}\right)
f(v)\,v\,dv\,d|\mathbf{q}|^2
\end{equation}
where the quantity $f$ is as defined in the main section.
Substituting from eqn.(\ref{dsig2}) in the
above equation we get
\begin{equation}
dR=\frac{\rho_\chi \sigma_p A^2}{4m_\chi m_N
\mu_p^2}F^2(E_R)\left(\frac{f(v)}{v}\right)dv\,d|\mathbf{q}|^2
\end{equation}

Using $d|\mathbf{q}|^2=2m_N\, dE_R$ we get the differential rate per
unit recoil energy,
\[
\frac{dR}{dE_R}=\frac{\rho_\chi\sigma_p
A^2}{2m_\chi\mu_p^2}F^2(E_R)\int_{v>v_{min}}\frac{f(v)}{v}dv
\]
which is same as eqn.(\ref{dsig}).


\begin{thebibliography}{99}

\bibitem{Drees}
M. Drees and G. Gerbier in PDG compilation, C. Amsler et al., Physics
Letters {\bf B667}, 1 (2008).

\bibitem {gju96}G.~Jungman, M.~Kamionkowski and K.~Griest, Phys. Rep.
\textbf{267,} 195 (1996).

\bibitem {xenon}J.~Angle \textit{et al. }[XENON10 Collaboration]\textit{,}
Phys. Rev. Lett. \textbf{101}, 091301 (2008)
[arXiv:astro-ph/0802.3530].

\bibitem {cdms}Z.~Ahmed \textit{et al. }[CDMS Collaboration]\textit{,}
Phys.Rev. Lett. \textbf{102}, 011301 (2009)
[arXiv:astro-ph/0802.3530].

\bibitem {dama}R.~Bernabei \textit{et al.} [DAMA Collaboration]\textit{,}
arXiv:0804.2741 [astro-ph].

\bibitem {cogent}C.~Aalseth \textit{et al.} [CoGeNT Collaboration]\textit{,}
Phys. Rev. Lett. \textbf{101}, 251301 (2008)
[arXiv:astro-ph/0807.0879].

\bibitem {gelmini}C.~Savage, G.~Gelmini, P.~Gondolo and K.~Freese,
arXiv:0808.3607 [astro-ph].

\bibitem{Bottino:2008mf}
A.~Bottino, F.~Donato, N.~Fornengo and S.~Scopel, Phys.\ Rev.\  D
{\bf 78}, 083520 (2008) [arXiv:hep-ph/0806.4099].

\bibitem{weiner}D.~Tucker-Smith and N.~Weiner, Phys. Rev. D {\bf 64},
043502
(2001) [arXiv:hep-ph/0101138];
D.~Tucker-Smith and N.~Weiner,
Phys.\ Rev.\  D {\bf 72}, 063509 (2005)
[arXiv:hep-ph/0402065];
S.~Chang, G.~D.~Kribs, D.~Tucker-Smith and N.~Weiner,
arXiv:0807.2250 [hep-ph].

\bibitem {mpos00}M.~Pospelov and T.~Veldhuis, Phys. Lett. \textbf{B
480}, 181
(2000) [arXiv:hep-ph/0003010].

\bibitem {ksig04}K.~Sigurdson, M.~Doran, A.~Kurylov, R.~R.~Caldwell
and M.~Kamionkowski,
  Phys.\ Rev.\  D {\bf 70}, 083501 (2004)
  [Erratum-ibid.\  D {\bf 73}, 089903 (2006)]
  [arXiv:astro-ph/0406355].

\bibitem{l3}
M.~Acciarri {\it et al.}  [L3 Collaboration],
e-
Phys.\ Lett.\  B {\bf 412}, 201 (1997).

\bibitem{wmap}
  E.~Komatsu {\it et al.}  [WMAP Collaboration], Astrophys.\ J.\
Suppl.\  {\bf 180}, 330 (2009)
  [arXiv:0803.0547 [astro-ph]].

\bibitem{barger}
V.~Barger, W.~Y.~Keung and G.~Shaughnessy,
Phys.\ Rev.\  D {\bf 78}, 056007 (2008)
[arXiv:0806.1962 [hep-ph]].


\bibitem{kolb}
  E.~W.~Kolb and M.~S.~Turner, Front.\ Phys.\  {\bf 69}, 1 (1990).

\bibitem{haber}
  H.~E.~Haber and G.~L.~Kane,
  Phys.\ Rept.\  {\bf 117}, 75 (1985).

\end{thebibliography}
\end{document}